\documentclass[10pt,letterpaper,english,reprint, aps]{revtex4-2}
\usepackage[T1]{fontenc}
\usepackage[utf8]{inputenc}
\usepackage{amsmath}
\usepackage{amssymb}
\usepackage{graphicx}



\usepackage{babel}
\begin{document}
\title{A Semiclassical Gaussian Wavepacket Method for Non-Adiabatic Molecular
Dynamics}
\author{Lorenzo Bocchi}
\affiliation{Dipartimento di Chimica, Università degli Studi di Milano, via Golgi
19, 20133 Milano, Italy}
\author{Jia-Xi Zeng}
\email{jiaxi.zeng@unimi.it}

\affiliation{Dipartimento di Chimica, Università degli Studi di Milano, via Golgi
19, 20133 Milano, Italy}
\author{Michele Ceotto}
\email{michele.ceotto@unimi.it}

\affiliation{Dipartimento di Chimica, Università degli Studi di Milano, via Golgi
19, 20133 Milano, Italy}
\date{\today}
\begin{abstract}
We introduce two non-adiabatic semiclassical methods that employ two
coupled Gaussian wavepackets, each one traveling on a separate diabatic
potential energy surface. The wavepackets take the form of thawed
Gaussians and are driven by classical equations of motion which account
for the diabatic coupling. The classical equations of motion are derived
in one case by enforcing the thawed Gaussian ansatz, while in the
other the time-dependent variational principles to the thawed Gaussian
ansatz. After a sanity check where both approximations reproduce Rabi
oscillations, the methods are applied to two non-adiabatic potential
energy scenarios. The first one involves two coupled displaced harmonic
oscillators, as in a typical electron transfer reaction. The second
one comprises a Morse potential coupled to an upper dissociative state,
modeling a photo-dissociation process. In both scenarios, the variational
thawed Gaussian approach is quite accurate, while the standard thawed
Gaussian one fails to fully capture the non-adiabatic effects. Ultimately,
non-adiabatic molecular dynamics is reproduced by means of two classical
trajectories without introducing any artificial jump or other \emph{ad-hoc}
non-classical effects.
\end{abstract}
\maketitle

\section{\label{sec:intro}Introduction}

Many important processes in nature involve non-adiabatic molecular
events, i.e. they can not be described within the Born-Oppenheimer
framework. Proton-coupled electron transfer reactions,\citep{schiffer2010PCETreview}
homogeneous and heterogeneous catalysis,\citep{homogeneous_catalysis2014,Benaglia_Ceotto2016kinetics,wodtke2016electronically,Fallacara_Ceotto_Fate_Formic_Acid_2024}
photochemical reactions,\citep{Garavelli_Mennucci2019review} charge
transport in materials\citep{Blumberger_chargetransport_22} and electron
transfer reactions\citep{Corni_electrontransfer_16} are just a few
examples of non-adiabatic processes in chemistry, physics and biology.

One can rigorously perform non-adiabatic dynamics and include both
nuclear and electronic quantum effects with exact grid-based methods,\citep{agostini2019WIRED}
as for example by employing multi-configurational time-dependent Hartree
methods,\citep{Meyer_Cederbaum_MCTDH_1990} or related methods, such
as variational multi-configurational Gaussian,\citep{GMCTDH2004}
full multiple spawning,\citep{fullMSpawning1997} and multi-configurational
Ehrenfest dynamics methods.\citep{shalashilin2009Ehrenfestexact,Doriol_cloning_17,Shalashilin_review_2017}
Alternatively, to alleviate the computational burden, a series of
mixed quantum-classical methods have been developed since the pioneering
surface hopping idea.\citep{tully_preston1971SH,Akimov_SH_25,barbatti2026newton,Barbatti_SH_25}
Spin-mapping methods\citep{Richardson_spinmapping_2019,Richardson_spinmappingSH_2023,Liu_nonadiabaticField_2025}
are an example of quantum-classical dynamics. Given the importance
of nuclear quantum effects, even in condensed phase processes,\citep{schleif_tunneling_solvation_2022,fausto_irMatrix_2022,mandelli_ceotto_solvation_2026}
methods based on classical trajectories and describing non-adiabatic
events have been developed. In this context, one can either employ
the Meyer-Miller Hamiltonian,\citep{Meyer_Miller_nonadiabatic_1979,Zeng_2025_scivr,Stock_Thoss_SC_nonadiabatic_1997,Church_Ananth_NonadiabSC_2018,moscatoceotto_nonadiabatic_2025,cottonmiller_faraday_2016,cottonmiller2017},
a frozen Gaussian basis in conjunction with the time-dependent variational
principle,\citep{Doriol_frozengaussians_18,Doriol_variational_22},
the phase-space electronic structure theory,\citep{subotnik_phasespacereview_26}
the path integral\citep{ananth2013mapping,nandini_faradayNOnadiabatic_2016,HuoPengfei2021RPMDnonad,WeiFang_SHPI_25}
or the Exact Factorization\citep{Agostini_Ibele_2026,Agostini_Ibele_PCCPreview_2024,Min_XF_2026}
formulations of non-adiabatic dynamics.

Here we focus on the simplest approach that one can employ for nuclear
quantum non-adiabatic molecular dynamics, which is the one based on
a single classical trajectory per electronic state. In the adiabatic
case, single trajectory approaches \citep{ceotto2009multiple,Ceotto_AspuruGuzik_PCCPFirstprinciples_2009,Sulc_Vanicek_CellularDephasing_2012,Wehrle_Vanicek_Oligothiophenes_2014,beguvsic_vanicek_otf_vibronic_2020}
showed that anharmonicity and quantum effects can be reproduced, at
least to some extent. Single-trajectory Ehrenfest dynamics is an example,\citep{Jiri_enhrenfest_21}
but it is mainly limited by the overestimation of the electronic coherence
because all electronic states share the same classical nuclear trajectory
and quantum nuclear effects are not accounted for. Very recently Vanicek's
group introduced a method which is based on a single trajectory and
does include nuclear quantum effects.\citep{Vanicek_TGAnonadiabatic_2026}
The method is an implementation of Heller's thawed Gaussian wavepacket
propagation,\citep{Heller_TdependentSC_1975,hagedorn1998,vanicek_family_2023}
where the wavepacket is driven by the classical evolution of its components.
In general, thawed Gaussian wavepacket dynamics (TGWD) is rather simple
to implement and it provides a clear interpretation of quantum dynamics,
because classical variables are intuitive, localized in space, and
only the potential values along the trajectory path are needed. In
addition, on-the-fly ab initio implementation is direct, as for many
semiclassical methods.\citep{Wehrle_Vanicek_NH3_2015,Wehrle_Vanicek_Oligothiophenes_2014,ceotto_conte_DCSCIVR_2017,Ceotto_Hase_AcceleratedSC_2013,Conte2019SemiclassicalDatabases,Conte_Ceotto_JPCLPerspective_2024,conte_Ceotto_perspectiveChemSci_2025,Gabas2017On-The-FlySpectrum,Gandolfi_Ceotto_NeuralGas_2021,Moscato_Ceotto_TimidinaJACS_2024}
The accuracy of TGWD can be improved by including the third derivative
of the potential, as in the case of the “extended” semiclassical wavepacket
dynamics\citep{Mitric_extendedTGA2023} and the symplectic semiclassical
wavepacket dynamics.\citep{ohsawa_TGAsymplectic_2013,Moghaddasi_Vanicek_localcubicdynamics_2024,pattanayak_extendedTGA_1994}

A less approximated approach but still based on a single trajectory
Gaussian wavepacket propagation is the Variational Thawed Gaussian
Wavepacket Dynamics (VTGWD), which is derived by enforcing the McLachlan
variational principle\citep{lubich_diracfrenkel2008} to the thawed
Gaussian ansatz.\citep{garashchuk2025variational} Already Heller,\citep{Heller_TDVP_1976}
and then Heather and Metiu\citep{Metiu_TDVP_1985} and Karplus and
Coalson\citep{Karplus_TDVP_1990} derived the equations of motion
for the wavepacket. In comparison to TGWD, VTGWD has the advantage
of being symplectic,\citep{lubich_diracfrenkel2008} energy preserving
and it can qualitatively describe tunneling.\citep{Moghaddasi_Vanicek_localcubicdynamics_2024,Vanicek_Rojia_VTGA_2023}
However, these accuracy improvements come at the cost of evaluating
the expectation values of the potential and its derivatives.

Despite the success of VTGWD in adiabatic scenarios, the extension
of this approach to non-adiabatic cases is still a challenge. In this
work, we tackle this issue by introducing a non-adiabatic method for
nuclear quantum dynamics based on the thawed Gaussian idea. The idea
is to have a wavepacket for each electronic potential energy surface
and have the potential coupling to induce the wavepacket population
to exchange.

The paper is organized as follows. In section \ref{sec:theory}, we
present the theoretical framework. The equations of motion of our
non-adiabatic thawed Gaussian wavepacket dynamics (NA-TGWD) are in
subsection \ref{subsec:A-Thawed-Gaussian}, while the ones of the
non-adiabatic variational thawed Gaussian wavepacket dynamics (NA-VTGWD)
are in subsection \ref{subsec:A-Variational-Gaussian}. The results
are presented in Section \ref{sec:Results}. In subsection \ref{subsec:Rabi-oscillations}
we perform a sanity check with a Rabi system. In subsection \ref{subsec:Electron-transfer-type}
we consider two coupled displaced harmonic oscillators, which describe
an electron-transfer type of reaction or a proton-electron transfer
reaction. In subsection \ref{subsec:Unimolecular-reaction-potential}
we present the results for a photodissociation or a non-adiabatic
unimolecular reaction potential energy profile. A Discussion and Conclusion
section (Sec. \ref{sec:Discussion-and-Conclusions}) concludes the
paper.

\section{\label{sec:theory}Semiclassical Wavepacket Non-Adiabatic Methods}

We formulate the equivalent of TGWD and VTGWD for the non-adiabatic
case by using the diabatic potential energy surface representation.
Specifically, we consider two coupled electronic states and calculate
the nuclear wavepacket evolution on both surfaces simultaneously with
the addition, with respect to the adiabatic case, that in our non-adiabatic
formulation the wavepackets are coupled. The idea is to reproduce
quantum electronic transitions by allowing the wavepacket shape to
change, without introducing any artificial hops or forcing classical
trajectories to change potential energy surface. Instead, we stick
with Hamilton equations of motion for the wavepacket center positions
and momenta. More specifically, modified Hamilton equations and equations
for the wavepacket widths and phases are derived in order to reproduce
as much as possible electronic transitions, as shown in Subsec.s \ref{subsec:A-Thawed-Gaussian}
and \ref{subsec:A-Variational-Gaussian}.

\subsection{\label{subsec:A-Thawed-Gaussian}A Thawed Gaussian Non-Adiabatic
Method}

The quantum nuclear evolution of two wavepackets on two coupled electronic
states in the diabatic framework is described by the set of equations
\begin{equation}
\begin{cases}
i\hbar\dfrac{\partial}{\partial t}\phi_{1}(x,t) & =-\dfrac{\hbar^{2}}{2m}\dfrac{\partial^{2}}{\partial x^{2}}\phi_{1}(x,t)+V_{11}(x)\phi_{1}(x,t)\\
 & +V_{12}(x)\phi_{2}(x,t)\\
i\hbar\dfrac{\partial}{\partial t}\phi_{2}(x,t) & =-\dfrac{\hbar^{2}}{2m}\dfrac{\partial^{2}}{\partial x^{2}}\phi_{2}(x,t)+V_{22}(x)\phi_{2}(x,t)\\
 & +V_{12}(x)\phi_{1}(x,t)
\end{cases}\label{tg}
\end{equation}
where $V_{11}(x)$ and $V_{22}(x)$ are the diabatic surfaces, $V_{12}(x)$
is the diabatic coupling, and $\phi_{i}\left(x,t\right)$ is a diabatic
wavepacket. In our semiclassical approach we adopt the thawed Gaussian
ansatz
\begin{align}
\phi_{1}(x,t) & =\left(\dfrac{2\alpha^{R}_{1}(0)}{\pi}\right)^{1/4}\exp\Big[\dfrac{i}{\hbar}S_{1}(t)+\dfrac{i}{\hbar}\varphi_{1}(t)\Big]\nonumber \\
\times & \exp\Big[-\alpha_{1}(t)\big(x-q_{1}(t)\big)^{2}+\dfrac{i}{\hbar}p_{1}(t)\big(x-q_{1}(t)\big)\Big]\label{eq:tg1_ansatz}\\
\phi_{2}(x,t) & =\left(\dfrac{2\alpha^{R}_{2}(0)}{\pi}\right)^{1/4}\exp\Big[\dfrac{i}{\hbar}S_{2}(t)+\dfrac{i}{\hbar}\varphi_{2}(t)\Big]\nonumber \\
\times & \exp\Big[-\alpha_{2}(t)\big(x-q_{2}(t)\big)^{2}+\dfrac{i}{\hbar}p_{2}(t)\big(x-q_{2}(t)\big)\Big]\label{eq:tg2_ansatz}
\end{align}
where $S_{i}(t)=\int^{t}_{0}\left(\dfrac{p^{2}_{i}(t')}{2m}-V_{i}(q_{i}(t'))\right)\,dt'$
is the usual classical action. The variables $\alpha_{i}(t)$ and
$\varphi_{i}(t)$ are complex-valued time-dependent quantities whose
equations of motion are to be determined together with the ones for
the wavepacket center $q_{i}\left(t\right)$ and momentum $p_{i}\left(t\right)$.
The real part $\alpha^{R}_{i}(t)$ is the width parameter of the wavepacket,
while the imaginary part represents a spatial chirp. The real part
of $\varphi_{i}(t)$ is a time-dependent phase factor which accounts
for the wavepacket quantum zero-point energy, while the imaginary
part ensures normalization at each time-step. Eq.s (\ref{eq:tg1_ansatz})
and (\ref{eq:tg2_ansatz}) are exact for constant, linear and harmonic
adiabatic potentials.

To solve Eq.s (\ref{tg}), we substitute Eq.s (\ref{eq:tg1_ansatz})
and (\ref{eq:tg2_ansatz}) and expand each function of the nuclear
coordinate $x$ around the respective wavepacket center $q_{i}\left(t\right)$.
The coupling term $V_{ij}\left(x\right)\phi_{j}$$\left(x,t\right)$
is rewritten as $V_{ij}\left(x\right)\left(\phi_{j}\left(x,t\right)/\phi_{i}\left(x,t\right)\right)\phi_{i}\left(x,t\right)$
and both the coupling potential and the ratio $\left(\phi_{j}\left(x,t\right)/\phi_{i}\left(x,t\right)\right)$
are expanded. The resulting coupled equations are solved by equating
the coefficients of the same polynomial $\big(x-q_{i}(t)\big)^{n}$
order. The entire procedure is reported in detail in Sec. IA of the
supplementary material. The resulting equations of motion for each
parameter driving the NA-TGWD are
\begin{align}
\dot{\alpha}_{i}(t) & =-\dfrac{2i\hbar}{m}\alpha^{2}_{i}(t)+\dfrac{i}{2\hbar}\left[V^{\prime\prime}_{ii}(q_{i})+V_{ij}(q_{i})\Big(\dfrac{\tilde{\phi}_{j}}{\tilde{\phi}_{i}}\Big)^{\prime\prime}\right.\nonumber \\
 & \left.+2V^{\prime}_{ij}(q_{i})\Big(\dfrac{\tilde{\phi}_{j}}{\tilde{\phi}_{i}}\Big)^{\prime}+V^{\prime\prime}_{ij}(q_{i})\Big(\dfrac{\tilde{\phi}_{j}}{\tilde{\phi}_{i}}\Big)\right],\label{eq:alpha_dot}
\end{align}
 for the wavepacket width, 
\begin{align}
\dot{q}_{i}(t) & =\dfrac{p_{i}(t)}{m}+\dfrac{1}{2\hbar\alpha^{R}_{i}(t)}\left[V_{ij}(q_{i}(t))\:\text{Im}\left(\dfrac{\tilde{\phi}_{j}}{\tilde{\phi}_{i}}\right)^{\prime}\right.\nonumber \\
 & \left.+V^{\prime}_{ij}(q_{i}(t))\:\text{Im}\left(\dfrac{\tilde{\phi}_{j}}{\tilde{\phi}_{i}}\right)\right]\label{eq:q_dot}\\
\dot{p}_{i}(t) & =-V^{\prime}_{ii}(t)-\left[V_{ij}(q_{i}(t))\text{Re}\left(\dfrac{\tilde{\phi}_{j}}{\tilde{\phi}_{i}}\right)^{\prime}+V^{\prime}_{ij}(q_{i}(t))\text{Re}\left(\dfrac{\tilde{\phi}_{j}}{\tilde{\phi}_{i}}\right)\right]\nonumber \\
 & -\dfrac{\alpha^{I}_{i}(t)}{\alpha^{R}_{i}(t)}\left[V_{ij}(q_{i}(t))\text{Re}\left(\dfrac{\tilde{\phi}_{j}}{\tilde{\phi}_{i}}\right)^{\prime}+V^{\prime}_{ij}(q_{i}(t))\text{Im}\left(\dfrac{\tilde{\phi}_{j}}{\tilde{\phi}_{i}}\right)\right],\label{eq:p_dot}
\end{align}
for determining the wavepacket center phase space evolution, and

\begin{align}
\dot{\varphi}_{i}(t) & =-\dfrac{\hbar^{2}}{m}\alpha_{i}(t)-V_{ij}(q_{i}(t))\dfrac{\tilde{\phi}_{j}}{\tilde{\phi}_{i}}\nonumber \\
+ & \dfrac{p_{i}(t)}{2\hbar\alpha^{R}_{i}(t)}\left[V_{ij}(q_{i}(t))\text{Im}\left(\dfrac{\tilde{\phi}_{j}}{\tilde{\phi}_{i}}\right)^{\prime}+V^{\prime}_{ij}(q_{i}(t))\text{Im}\left(\dfrac{\tilde{\phi}_{j}}{\tilde{\phi}_{i}}\right)\right]\label{eq:phi_dot}
\end{align}
for the wavepacket phase. In Eq.s (\ref{eq:alpha_dot}), (\ref{eq:q_dot}),
(\ref{eq:p_dot}) and (\ref{eq:phi_dot}) $\tilde{\phi}_{k}=\phi_{k}\left(x=q_{i}\left(t\right),t\right)$
where $q_{i}\left(t\right)$ is the classical trajectory position
of the $i$-th diabatic state. Also, $\text{Re}\left(\phi_{j}/\phi_{i}\right)$
and $\alpha^{R}_{i}(t)$ are the real part, and $\text{Im}\left(\phi_{j}/\phi_{i}\right)$
and $\alpha^{I}_{i}(t)$ the imaginary part of the respective quantities.
Of course, we recover the original adiabatic thawed Gaussian equations
of motion when we take $i=j$ and $V_{12}=0$ in all the equations
(\ref{eq:alpha_dot})-(\ref{eq:phi_dot}). In other words, the non-adiabatic
motion is composed of additive terms: One is the same as in the single
surface adiabatic case, while the others account for the non-adiabatic
coupling. Unfortunately, Eq.s (\ref{eq:alpha_dot})-(\ref{eq:phi_dot})
do not conserve the norm, $N\left(t\right)=\int^{+\infty}_{-\infty}\left(|\phi_{1}\left(x,t\right)|^{2}+|\phi_{2}\left(x,t\right)|^{2}\right)dx$,
under time-evolution. This is an important limitation and we provide
a proof in Sec. I B of the supplementary material.

A more accurate single trajectory Gaussian wavepacket method is the
VTGWD, as anticipated in the Introduction. VTGWD is more accurate
because the equations of motion are obtained from the time-dependent
variational principle. This approach guarantees that at each time-step
each wavepacket variable is best fitted to the solution of the Schrödinger
equation. One can prove, besides being more accurate, that VTGWD is
symplectic, it exactly conserves energy and norm, and it is time-reversible.

\subsection{\label{subsec:A-Variational-Gaussian}A Variational Gaussian Non-Adiabatic
Method}

The application of the McLachlan time-dependent variational principle\citep{lubich_diracfrenkel2008,Doriol_variational_22,Karplus_TDVP_1990}
to the set of Eq.s (\ref{tg}) yields

\begin{equation}
\begin{cases}
\delta\int^{+\infty}_{-\infty}dx\,\left|i\hbar\partial_{t}\phi_{1}-(\hat{T}\phi_{1}+V_{11}\phi_{1}+V_{12}\phi_{2})\right|^{2} & =0\\
\delta\int^{+\infty}_{-\infty}dx\,\left|i\hbar\partial_{t}\phi_{2}-(\hat{T}\phi_{2}+V_{22}\phi_{2}+V_{12}\phi_{1})\right|^{2} & =0
\end{cases}\label{eq:McLachlan_equations}
\end{equation}
By introducing the collective variable $\varTheta_{i}\equiv\left(q_{i}\left(t\right),p_{i}\left(t\right),\alpha_{i}\left(t\right),\varphi_{i}\left(t\right)\right)=\left(\theta_{i1},...,\theta_{i4}\right)$,
the first set of Eq.s (\ref{eq:McLachlan_equations}) is equivalent
to
\begin{align}
\int^{+\infty}_{-\infty}dx\,\frac{\partial}{\partial\dot{\theta}^{*}_{1j}}\left|i\hbar\partial_{t}\phi_{1}-(\hat{T}\phi_{1}+V_{11}\phi_{1}+V_{12}\phi_{2})\right|^{2} & =\nonumber \\
-i\hbar\int dx\left(\frac{\partial\phi_{1}}{\partial\theta_{1j}}\right)^{*}\left[\left(i\hbar\partial_{t}-\hat{T}+V_{11}\right)\phi_{1}+V_{12}\phi_{2}\right] & =0\label{eq:variational_generic}
\end{align}
and a similar one is valid for the second set. This is described in
details in Sec. II A of the supplementary material. The single terms
in Eq. (\ref{eq:variational_generic}) can be evaluated as shown in
Sec. II B, II C and II D of the supplementary material. Setting up
and solving the systems of equations obtained from (\ref{eq:variational_generic})
and using the ansatz of Eq.s (\ref{eq:tg1_ansatz}) and (\ref{eq:tg2_ansatz})
where the classical action $S_{i}\left(t\right)$ has been incorporated
into the corresponding phase $\varphi_{i}\left(t\right)$, yields
the following equations of motion for the NA-VTGWD
\begin{equation}
\dot{q}_{1}=\frac{p_{1}}{m}+\frac{\text{Im}(C^{(1)}_{1})}{2\hbar\alpha^{R}_{1}}\label{eq:qdot}
\end{equation}
\begin{equation}
\dot{p}_{1}=-\text{Re}(C^{(1)}_{1})-\text{Im}(C^{(1)}_{1})\frac{\alpha^{I}_{1}}{\alpha^{R}_{1}}\label{eq:pdot}
\end{equation}

\begin{subequations}
\begin{align}
\dot{\alpha}_{1} & = & -\frac{2i\hbar}{m}\alpha^{2}_{1}+\frac{i}{2\hbar}\left(\frac{2\alpha^{R}_{1}}{\pi}\right)^{1/2}\langle V^{\prime\prime}_{11}\rangle_{11G}\label{eq:alpha0}\\
 &  & +\frac{i}{2\hbar}\frac{16(\alpha^{R}_{1})^{2}}{P_{1}(t)}\left[\langle\xi^{2}_{1}V_{12}\rangle_{12}-\frac{\langle V_{12}\rangle_{12}}{4\alpha^{R}_{1}}\right]\label{eq:alpha1}
\end{align}
\end{subequations}

\begin{subequations}
\begin{align}
\dot{\varphi}_{1} & = & \frac{p^{2}_{1}}{2m}-\left(\frac{2\alpha^{R}_{1}}{\pi}\right)^{1/2}\langle V_{11}\rangle_{11G}-\frac{\hbar^{2}}{m}\alpha_{1}\label{eq:phidot0}\\
 &  & +\left(\frac{2\alpha^{R}_{1}}{\pi}\right)^{1/2}\frac{\langle V^{\prime\prime}_{11}\rangle_{11G}}{8\alpha^{R}_{1}}\label{eq:phidot1}\\
 &  & +\frac{p_{1}\text{Im}(C^{(1)}_{1})}{2\hbar\alpha^{R}_{1}}\label{eq:phidot2}\\
 &  & +\frac{1}{P_{1}(t)}\left[-\frac{3}{2}\langle V_{12}\rangle_{12}+2\alpha^{R}_{1}\langle\xi^{2}_{1}V_{12}\rangle_{12}\right]\label{eq:phidot3}
\end{align}
\end{subequations}

where we omitted the time for each variable, and

\begin{subequations}
\begin{align}
C^{(1)}_{1} & =\left(\frac{2\alpha^{R}_{1}}{\pi}\right)^{1/2}\langle V^{\prime}_{11}\rangle_{11G}\label{eq:C1line1}\\
 & +\frac{4\alpha^{R}_{1}}{P_{1}(t)}\left\{ \frac{\langle V^{\prime}_{12}\rangle_{12}}{2G^{(2)}_{12}}+\left[\frac{G^{(1)}_{12}}{2G^{(2)}_{12}}-q_{1}\right]\langle V_{12}\rangle_{12}\right\} \label{eq:C1line2}
\end{align}
\end{subequations}

\begin{equation}
\langle V{}^{\prime\prime}_{11}\rangle_{11G}=\int^{+\infty}_{-\infty}V{}^{\prime\prime}_{11}(x)e^{-2\alpha^{R}_{1}(t)\left(x-q_{1}\left(t\right)\right){}^{2}}dx\label{eq:V11G}
\end{equation}
\begin{equation}
\langle V^{n}_{12}\rangle_{12}=\int^{+\infty}_{-\infty}V^{n}_{12}(x)\phi^{*}_{1}(x,t)\phi_{2}(x,t)\:dx\label{eq:V12}
\end{equation}

\begin{subequations}
\begin{align}
\langle\xi^{2}_{1}V_{12}\rangle_{12} & =\frac{\langle V^{\prime\prime}_{12}\rangle_{12}}{4(G^{(2)}_{12})^{2}}+\frac{\langle V^{\prime}_{12}\rangle_{12}}{2(G^{(2)}_{12})^{2}}\left[G^{(1)}_{12}-2q_{1}G^{(2)}_{12}\right]\label{eq:ChiV12a}\\
 & +\left[\left(\frac{G^{(1)}_{12}}{2G^{(2)}_{12}}-q_{1}\right)^{2}+\frac{1}{2G^{(2)}_{12}}\right]\langle V_{12}\rangle_{12}\label{eq:ChiV12b}
\end{align}
\end{subequations}

\begin{equation}
P_{1}(t)=\int^{+\infty}_{-\infty}|\phi_{1}(x,t)|^{2}dx=\left(\frac{\alpha^{R}_{1}(0)}{\alpha^{R}_{1}(t)}\right)^{1/2}e^{-\frac{2}{\hbar}\varphi^{I}_{11}(t)}\label{eq:P(t)}
\end{equation}

\begin{equation}
G^{(1)}_{12}=2(\alpha^{*}_{1}q_{1}+\alpha_{2}q_{2})+\frac{i}{\hbar}(p_{2}-p_{1})\label{eq:G12(1)}
\end{equation}
\begin{align}
G^{(2)}_{12} & =\alpha^{*}_{1}+\alpha_{2}\label{eq:G12(2)}
\end{align}
\begin{equation}
G^{(0)}_{12}=-(\alpha^{*}_{1}q^{2}_{1}+\alpha_{2}q^{2}_{2})+\frac{i}{\hbar}(p_{1}q_{1}-p_{2}q_{2})+\frac{i}{\hbar}(\varphi_{2}-\varphi^{*}_{1})\label{eq:G12(0)}
\end{equation}
Eq.s (\ref{eq:qdot})-(\ref{eq:phidot3}) provide some interesting
physical insights on how the non-adiabatic processes can be reproduced
with a pair of coupled classical trajectories and by also taking into
account quantum nuclear delocalization. Specifically, Eq. (\ref{eq:qdot})
for the position evolution is composed, as in the thawed Gaussian
case, of the usual adiabatic velocity plus a ``frictional'' contribution
in Eq. (\ref{eq:C1line2}) which accounts for the diabatic coupling.
Similarly, for the momentum evolution in Eq. (\ref{eq:pdot}), the
term in Eq. (\ref{eq:C1line1}) is the force averaged over the wavepacket
distribution, and the other term in Eq. (\ref{eq:C1line2}) describes
the non-adiabatic effects. The wavepacket width parameter is also
evolving as a direct sum of an adiabatic and non-adiabatic term. The
adiabatic one is given by Eq. (\ref{eq:alpha0}) and the non-adiabatic
one by Eq. (\ref{eq:alpha1}). Regarding the wavepacket phase in Eq.s
(\ref{eq:phidot0})-(\ref{eq:phidot3}), we recognize the Lagrangian
in Eq. (\ref{eq:phidot0}) plus a term which depends on the width
$\alpha_{1}\left(t\right)$ evolution. The terms in line (\ref{eq:phidot0})
and line (\ref{eq:phidot1}) are the same as in the adiabatic case.
An interesting term is the one on line (\ref{eq:phidot1}), which
reproduces the wavepacket quantum delocalization. The non-adiabatic
phase $\varphi_{1}\left(t\right)$ contribution is given by lines
(\ref{eq:phidot2}) and (\ref{eq:phidot3}). Interestingly, line (\ref{eq:phidot2})
takes the form of $p_{1}$ times the non-adiabatic velocity contribution,
while line (\ref{eq:phidot3}) represents the coupling potential $\left\langle V_{12}\right\rangle _{12}$
between the two wavepackets and its quantum delocalization $\langle\xi^{2}_{1}V_{12}\rangle_{12}$.
From these observations we conclude that the term $C^{(1)}_{1}$ is
pivotal for non-adiabatic coupling and it is composed of a real part,
the quantum delocalization in line (\ref{eq:C1line1}), and an imaginary
part, the non-adiabatic coupling in line (\ref{eq:C1line2}). In conclusion,
as in NA-TGWD, the non-adiabatic motion in NA-VTGWD is composed of
additive terms which are either adiabatic, i.e. the same as in the
single surface case, or non-adiabatic. Within this formulation, one
can also appreciate and control the amount of non-adiabaticity.

In Sec. II E of the supplementary material we look at the norm conservation,
as in the thawed Gaussian case, and prove that the norm is always
conserved for NA-VTGWD of Eq.s (\ref{eq:qdot})-(\ref{eq:G12(2)}).

\section{\label{sec:Results}Results}

We start our simulations with a sanity check of the two methods and
reproduce Rabi oscillations. Then, we move to more realistic scenarios
and simulate an electron transfer and a photodissociation population
inversion.

\subsection{Rabi oscillations\label{subsec:Rabi-oscillations}}

In the Rabi system, both diagonal and off-diagonal diabatic potential
terms are constant. The potential terms in Eq.s (\ref{eq:alpha_dot},
\ref{eq:q_dot}, \ref{eq:p_dot}, \ref{eq:phi_dot}) for NA-TGWD and
in Eq.s (\ref{eq:qdot}, \ref{eq:pdot}, \ref{eq:alpha0}, \ref{eq:alpha1},
\ref{eq:phidot0}, \ref{eq:phidot1}, \ref{eq:phidot2}, \ref{eq:phidot3})
for NA-VTGWD can be calculated analytically. Only the time-evolution
is performed numerically using a fourth-order Runge-Kutta algorithm.
We are aware that better algorithms are available for TGWD and especially
for VTGWD,\citep{Vanicek_Rojia_VTGA_2023} however for our purposes
we found the standard Runge-Kutta algorithm to be accurate enough.
The exact results are obtained by solving Eq.s (\ref{tg}) using the
split-operator quantum time-evolution\citep{split-operator} of the
same initial Gaussian wavepackets employed for NA-TGWD and NA-VTGWD.
Specifically, the Rabi model potential is
\begin{equation}
V=\begin{pmatrix}2.0 & 1.0\\
1.0 & 1.0
\end{pmatrix}\label{eq:Rabi_pot}
\end{equation}
and the wavepackets are evolving on completely flat potentials which
are coupled to each other by a constant coupling. In this case the
potential elements for the wavepacket time-evolution reduce to the
following:
\begin{align}
\langle V_{ii}\rangle_{iiG} & =V_{ii}\left(\frac{\pi}{2\alpha^{R}_{i}(t)}\right)^{1/2}\\[6pt]
\langle V_{12}\rangle_{12} & =V_{12}\left(\frac{4\alpha^{R}_{1}(0)\alpha^{R}_{2}(0)}{(G^{(2)}_{12})^{2}}\right)^{1/4}\exp\left[\frac{(G^{(1)}_{12})^{2}}{4G^{(2)}_{12}}+G^{(0)}_{12}\right]
\end{align}
The simulation was performed on a spatial grid ranging from $x_{\text{min}}=-20.0$
a.u. to $x_{\text{max}}=20.0$ a.u. using $N_{g}=1024$ grid points.
The temporal evolution was propagated for a total time of $10.0$
a.u., using a time-step of $dt=0.01$ a.u. (yielding 1000 total steps).
The masses are set to $m=1.0$ a.u.. Electronic states are initialized
with identical Gaussian wavepackets, i.e. the same shape and initial
population. The initial population $P_{i}\left(0\right)$ is specified
for each wavepacket and the imaginary part of the initial phase is
determined by inverting the population definition, $P_{i}\left(0\right)=\text{exp}\left[-2\varphi^{I}_{i}\left(0\right)\right]$.
In all our simulations, the real part of the initial phase is set
to zero.

\begin{figure}
\begin{centering}
\includegraphics[scale=0.3]{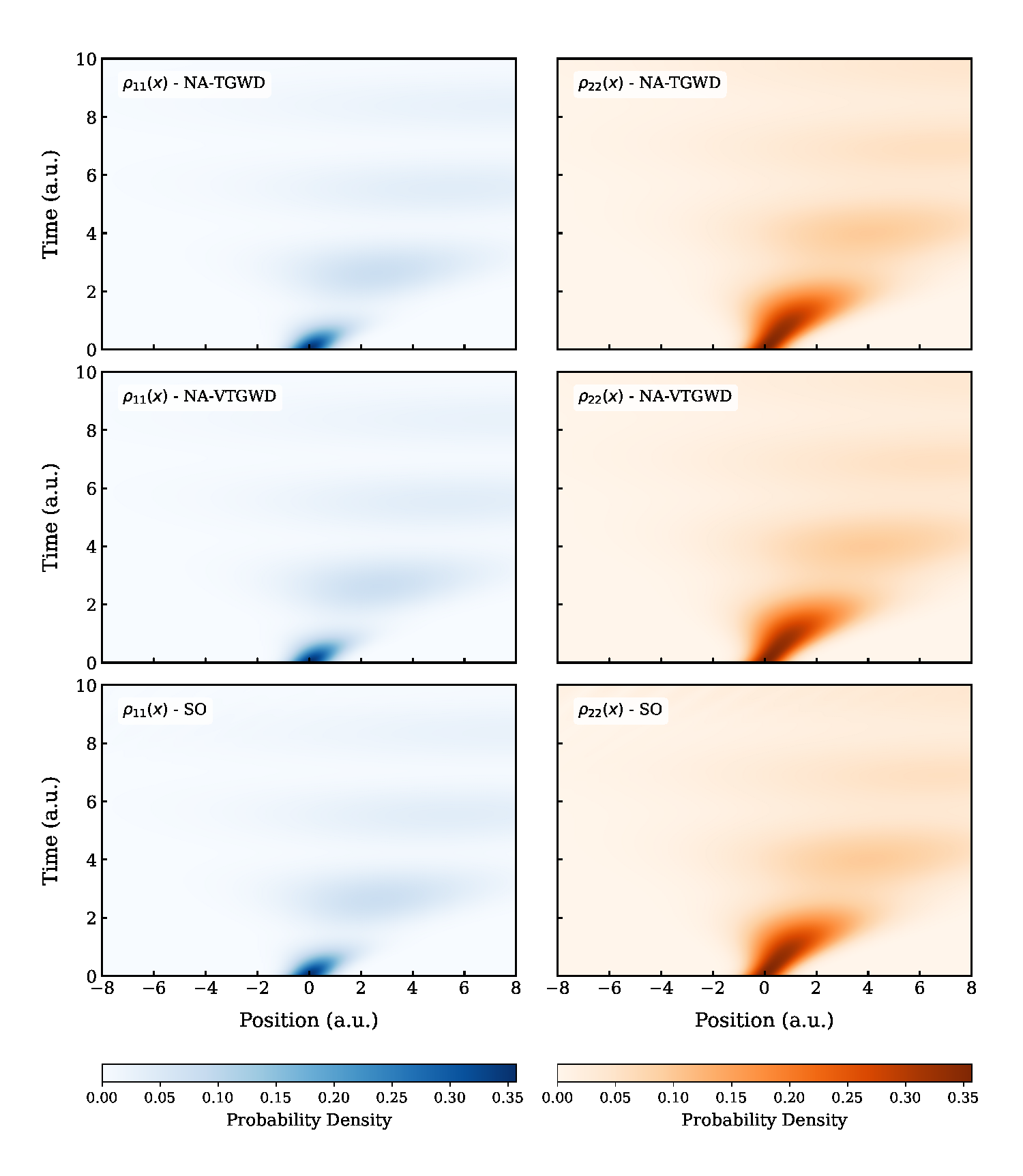}
\par\end{centering}
\caption{\label{fig:Rabi_heatmaps}Wavepacket heatmaps for the probability
time-evolution for the Rabi system of Eq. (\ref{eq:Rabi_pot}). Left
for the upper $V_{11}$ electronic state. Right for the lower $V_{22}$
one. Bottom panel is the exact quantum evolution, middle one is NA-VTGWD,
while upper panel is for NA-TGWD approximation.}
\end{figure}

\begin{figure}
\begin{centering}
\includegraphics[scale=0.3]{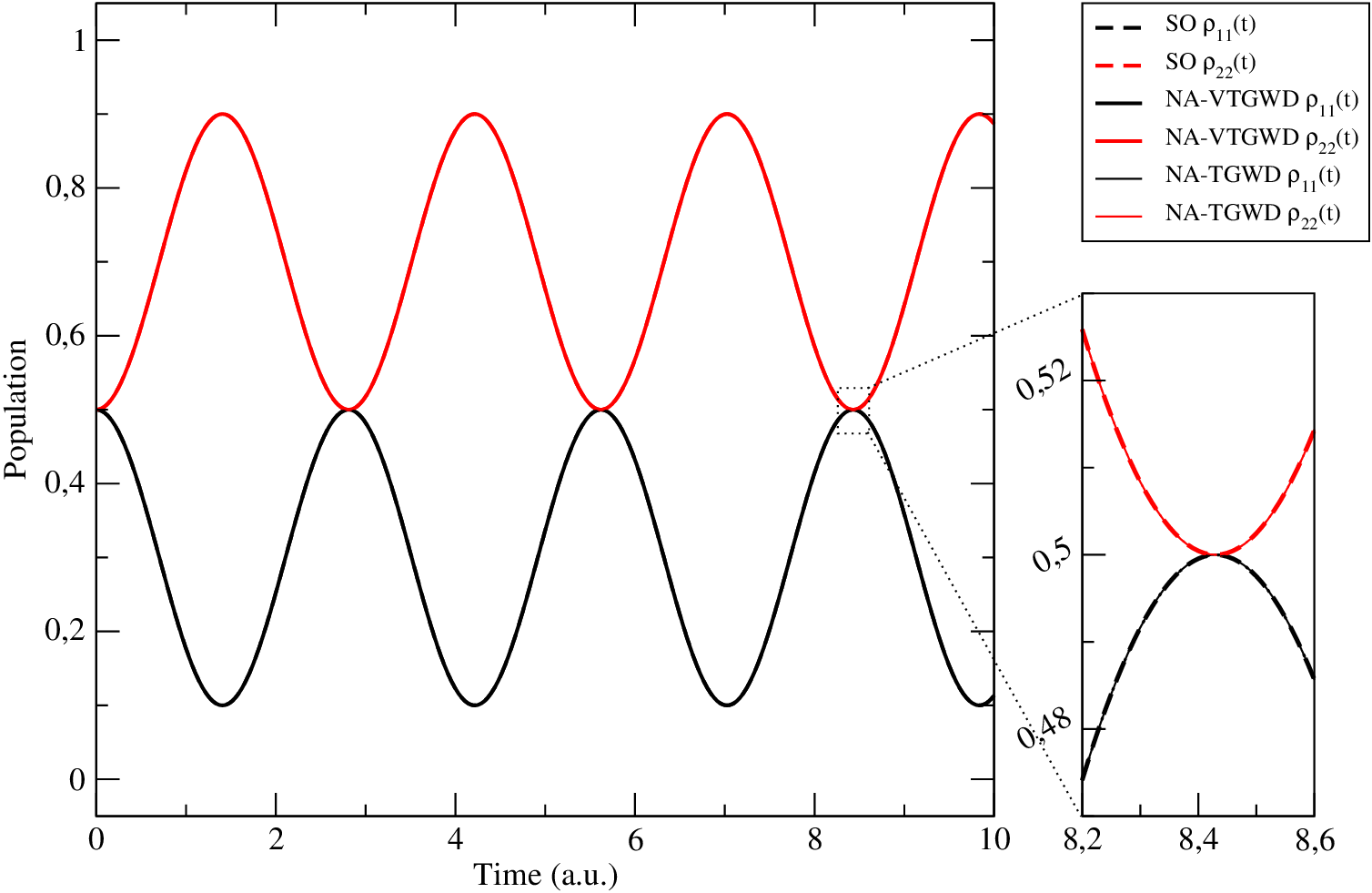}
\par\end{centering}
\caption{\label{fig:Rabi_populations}Wavepacket populations. Dashed lines
are for the exact split-operator wavepacket propagation. Thick solid
lines are for NA-VTGWD method, while thin solid lines for NA-TGWD
approximation. Inset shows the accuracy of the methods.}
\end{figure}

Fig. (\ref{fig:Rabi_heatmaps}) and Fig. (\ref{fig:Rabi_populations})
show respectively the wavepacket density heatmaps and the time-dependent
diabatic state populations. In both cases, NA-TGWD and NA-VTGWD simulate
the exact dynamics, and Rabi oscillations are exactly reproduced in
Fig. (\ref{fig:Rabi_populations}). We believe that the very small
deviations that one can appreciate only in the zooming panel of Fig.
(\ref{fig:Rabi_populations}) are due to the numerical algorithm integration,
which is confirmed to be accurate enough for our purpose.

As a second sanity check, we consider two coupled harmonic electronic
states where two wavepackets are oscillating with opposite phases,
i.e. the wavepacket on one surface is traveling in the opposite direction
to the one on the other surface. The two states are identical, as
are the potential energy surfaces. Thus, if the numerical integration
is accurate, the population is invariant with respect to the time-evolution.
Also in this case the analytical solution is accurately reproduced
both by NA-TGWD and NA-VTGWD, as reported in Fig. S1 and Fig. S2,
where the initial population is equally distributed between the two
states.

After checking that both methods accurately reproduce basic non-adiabatic
dynamics, we can proceed to test them on more complex potential energy
profiles.

\subsection{\label{subsec:Electron-transfer-type}Electron transfer type of reactions}

\begin{figure*}
\begin{centering}
\includegraphics[scale=0.5]{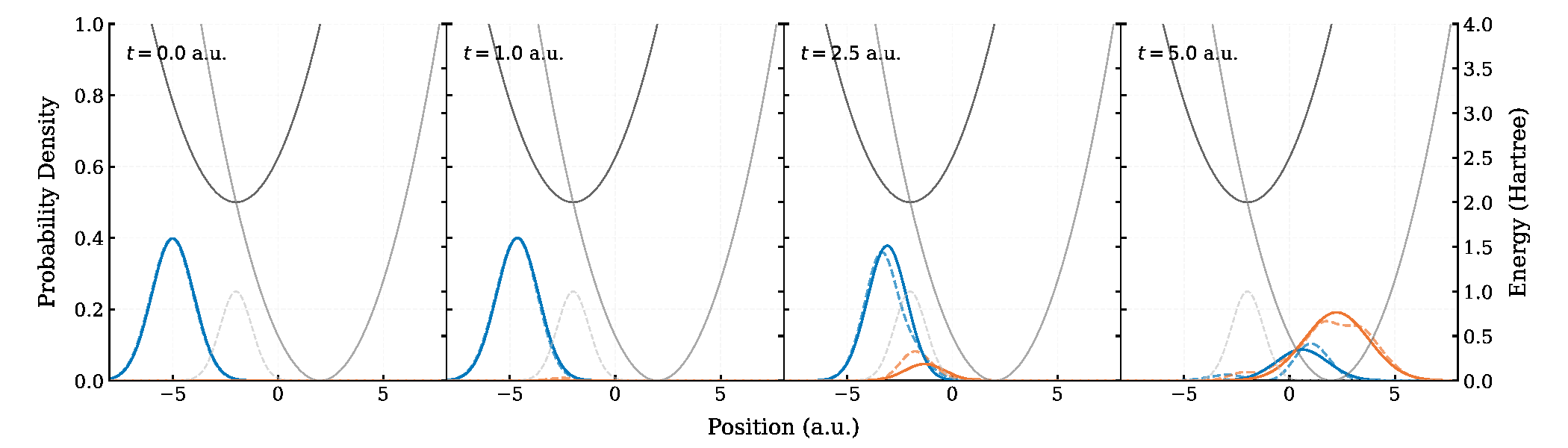}
\par\end{centering}
\caption{\label{fig:dispHOs_wavepackets}Diabatic wavepackets time-evolution.
Gray for the potential terms, blue for the upper $V_{11}$ state wavepacket
and orange for the lower $V_{22}$ state one. Colored solid lines
for NA-VTGWD, and colored dashed lines for the exact split-operator
quantum time evolution.}
\end{figure*}

A more challenging system for non-adiabatic molecular dynamics is
that one reported in Fig. (\ref{fig:dispHOs_wavepackets}) by light
and dark gray lines. These types of diabatic potentials reproduce,
at least locally, the shape of a typical electronic structure crossing
and they serve as model potentials for electron-transfer reactions.
Specifically, the potential energy surface is given by
\begin{equation}
\begin{pmatrix}\dfrac{1}{2}m\omega^{2}(x-x^{\text{eq}}_{11})^{2}+2.0 & Ce^{-D(x-x^{\text{eq}}_{12})^{2}}\\[8pt]
Ce^{-D(x-x^{\text{eq}}_{12})^{2}} & \dfrac{1}{2}m\omega^{2}(x-x^{\text{eq}}_{22})^{2}
\end{pmatrix}\label{eq:PES}
\end{equation}
where we set the mass to $m=1.0$ a.u. and the harmonic frequency
to $\omega=0.5$ a.u.. The equilibrium positions are located at $x^{\text{eq}}_{11}=-2.0$
a.u. for the upper state and $x^{\text{eq}}_{22}=2.0$ a.u. for the
lower state. The $V_{12}$ term is given by a Gaussian-type coupling
centered at $x^{\text{eq}}_{12}=-2.0$ a.u.. We intentionally placed
the coupling at the minimum of the upper potential to have the wavepackets
interacting from the very beginning of the dynamics. We choose the
coupling constant to be $C=1.0$ a.u. and $D=0.888$ a.u.. The wavepacket
initial positions are $q_{1}\left(0\right)=-5.0\:\text{a.u.}$ and
$q_{2}\left(0\right)=0.0\:\text{a.u.}$, the initial momenta are both
zero, the complex Gaussian widths are the same for both wavepackets,
with $\alpha_{i}\left(t\right)=\left(0.250,0.000\right)\:\text{a.u.}$,
and the initial populations are respectively 0.999 and 0.001. The
empty state population is not exactly zero to avoid numerical issues
in the coupling term calculation. The wavepackets are evolved with
a time-step of 0.01 a.u. for a total of 1000 steps. The split-operator
simulation is performed within a grid span of $\pm20.0$ a.u. sampled
by 1024 equally spaced grid-points and with a Gaussian wavepacket
width parameter equal to that of the initial thawed Gaussian wavepackets
for both surfaces.

 Fig. (\ref{fig:dispHOs_wavepackets}) reports four snapshots of the
wavepacket time-evolution and the diabatic potential terms described
above as a reference. NA-VTGWD is reported with colored solid lines:
Blue for the upper state wavepacket and orange for the lower state
one. The exact split-operator wavepacket is reported in Fig. (\ref{fig:dispHOs_wavepackets})
with the same color code but using dashed lines. These snapshots allow
us to appreciate how NA-VTGWD fits the shape of the exact wavepacket
with a Gaussian one at each time-step.

\begin{figure}
\begin{centering}
\includegraphics[scale=0.3]{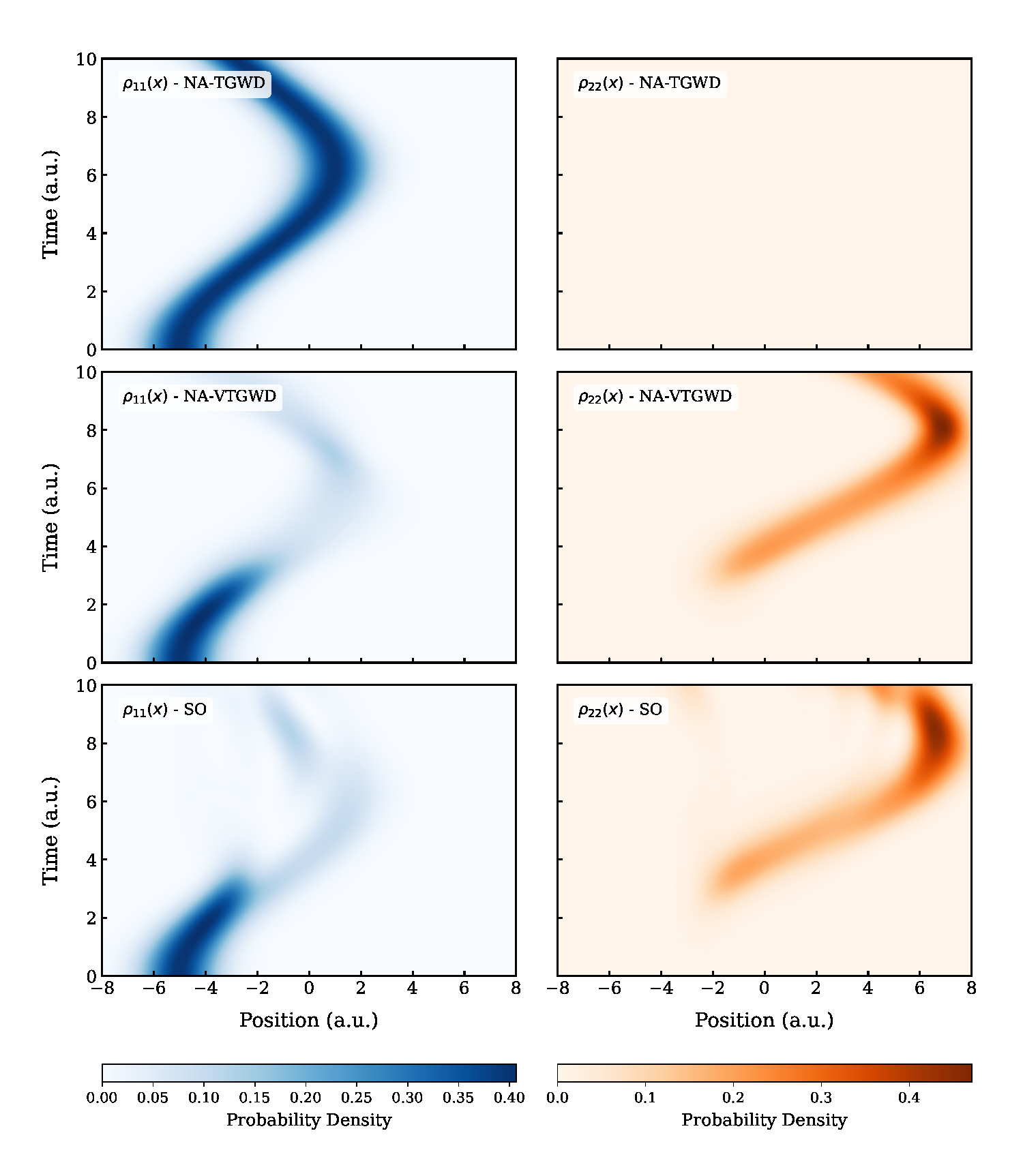}
\par\end{centering}
\caption{\label{fig:dispHOs_heatmaps}Wavepacket heatmaps. Left for the upper
$V_{11}$ electronic state. Right for the lower $V_{22}$ one. Bottom
panel is the exact quantum evolution, middle one is NA-VTGWD, while
upper panel is for NA-TGWD approximation.}
\end{figure}
 Fig. (\ref{fig:dispHOs_heatmaps}) reports a heatmap representation
of the wavepacket dynamics. In this way one can appreciate the accuracy
of each method at each time-step, as well as its ability to reproduce
quantum delocalization. While NA-TGWD completely misses the non-adiabatic
effects, as reported on the upper panels of Fig. (\ref{fig:dispHOs_heatmaps}),
the NA-VTGWD results in the middle panels (left panel for the upper
state and right one for the lower state) mimic quite well the exact
split-operator maps of the bottom panels. What NA-VTGWD can not capture
is the interference pattern generated by the wavepacket splitting
on the same potential energy surface. This is expected since the NA-VTGWD
ansatz is forced to always be a single Gaussian wavepacket. However,
since the wavepacket splitting within the same surface is limited,
the single Gaussian ansatz effectively handles this quantum nuclear
delocalization.

\begin{figure}
\begin{centering}
\includegraphics[scale=0.3]{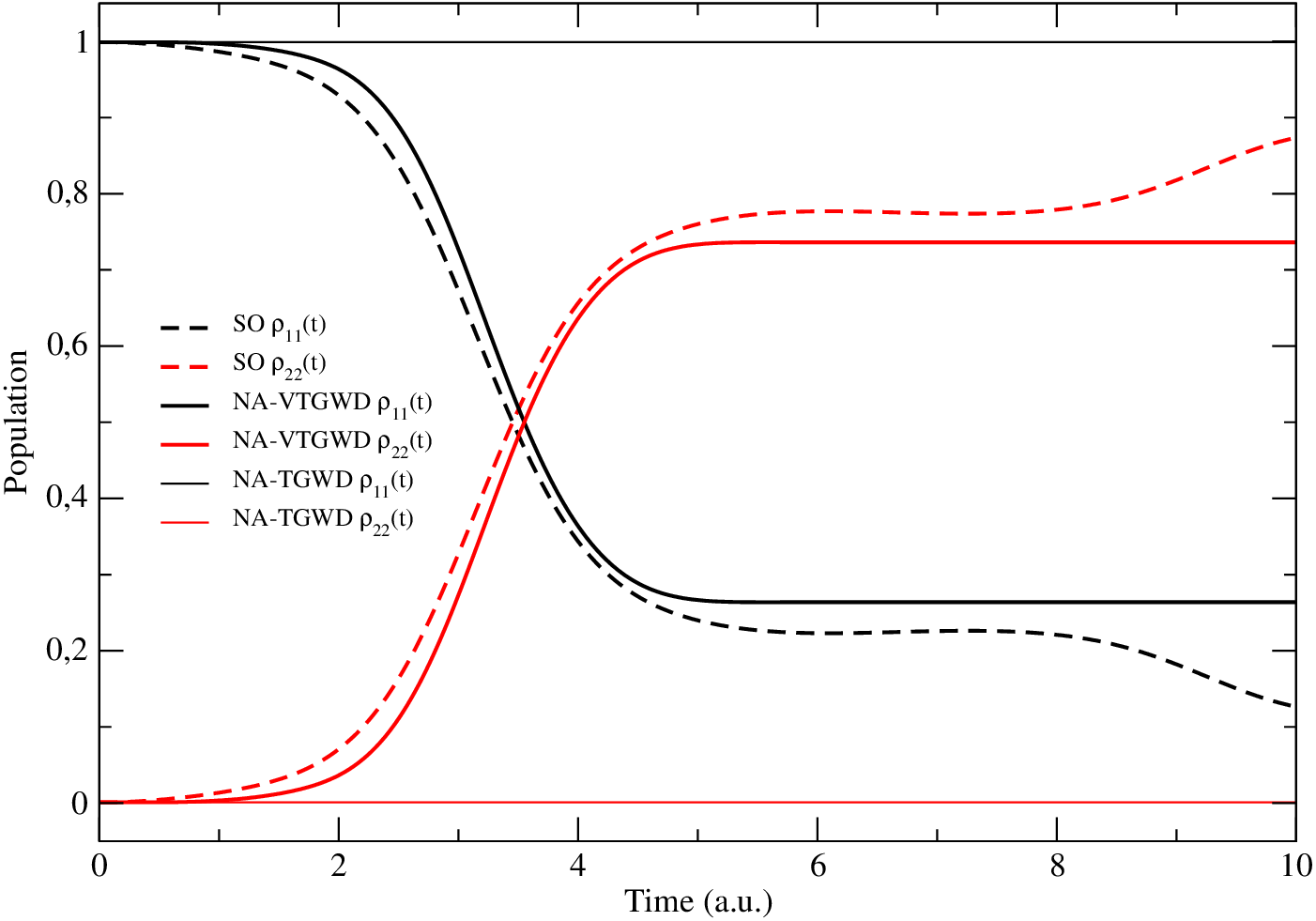}
\par\end{centering}
\caption{\label{fig:dispHOs_populations}Wavepacket populations. Dashed lines
are for the exact split-operator wavepacket propagation. Thick solid
lines are for NA-VTGWD method, while thin solid lines for NA-TGWD
approximation.}
\end{figure}

To quantify the accuracy of both methods we report in Fig. (\ref{fig:dispHOs_populations})
the population of each electronic state for each time-step.
 Specifically, the thin black and red lines report the population
calculated by the NA-TGWD method. As anticipated by the heatmaps in
Fig. (\ref{fig:dispHOs_heatmaps}), in this case there is no population
exchange despite the local diabatic coupling. Instead, the thick red
and black lines show the population inversion between the electronic
states for the NA-VTGWD method, which mimics quite well the exact
populations reported by colored dashed lines.

\subsection{\label{subsec:Unimolecular-reaction-potential}Photodissociation
reaction potential energy profile}

We now consider a potential energy profile which reproduces a photodissociation-type
reaction, where a ground state is photoexcited to an unbound upper
electronic state and the wavepacket decays into a dissociative configuration.
The potential is given by
\begin{equation}
\begin{pmatrix}V_{11}(x) & Ce^{-D(x-x^{\text{eq}}_{12})^{2}}\\[8pt]
Ce^{-D(x-x^{\text{eq}}_{12})^{2}} & Ee^{-F(x-x^{\text{eq}}_{22})}+0.5
\end{pmatrix}\label{eq:PES2}
\end{equation}
where $V_{11}(x)$ represents the bound Morse potential $V_{11}\left(x\right)=D_{e}\left(1-\text{exp}\left[-b\left(x-x^{eq}_{11}\right)\right]\right)^{2}$.
We present two different cases: A deep Morse potential, with parameters
$D_{e}=15.0$ a.u. and $b=0.091287$ a.u., and a shallow Morse potential,
with $D_{e}=4.0$ a.u. and $b=0.17678$ a.u.. In both cases we choose
$x^{\text{eq}}_{11}=-2.0$ a.u.. The dissociative potential parameters
are $E=0.1$ a.u., $F=0.4$ a.u., and $x^{\text{eq}}_{22}=8.0$ a.u..
The coupling potential parameters are $C=1.5$ a.u., $D=0.888$ a.u.
in both cases, and the coupling is centered at $x^{\text{eq}}_{12}=2.195\;\text{a.u.}$
for the deep Morse case, and $x^{\text{eq}}_{12}=2.79\;\text{a.u.}$
for the shallow Morse case. The split-operator simulation is performed
starting with a Gaussian wavepacket over a grid extension of $\pm40.0$
a.u. with 2048 equally spaced grid-points, and the wavepacket evolution
employs a time-step of 0.01 a.u. for a total of 1000 steps. The initial
Gaussian width is set at $\alpha_{i}\left(0\right)=\left(0.250,0.000\right)\:\text{a.u.}$
for all wavepackets. The photoexcited wavepacket, with a population
of 0.999, starts at $q_{2}\left(0\right)=-2.00\:\text{a.u.}$ with
an initial momentum $p_{2}\left(0\right)=0.0\:\text{a.u.}$, and it
has enough initial energy to dissociate along the non-adiabatic path,
i.e. along the Morse potential profile, only for the shallow Morse
case. The bound Morse state wavepacket has a very small initial population
of 0.001 and in both cases it is initialized such that it crosses
the coupling region at the same time as the other wavepacket. Specifically,
for the deep Morse case it starts at the phase space point $q_{1}\left(0\right)=-5.00\:\text{a.u.}$
and $p_{1}\left(0\right)=3.00\:\text{a.u.}$, while for the shallow
Morse case it starts at $q_{1}\left(0\right)=-5.14\:\text{a.u.}$
and $p_{1}\left(0\right)=2.74\:\text{a.u.}$

\begin{figure*}
\begin{centering}
\includegraphics[scale=0.5]{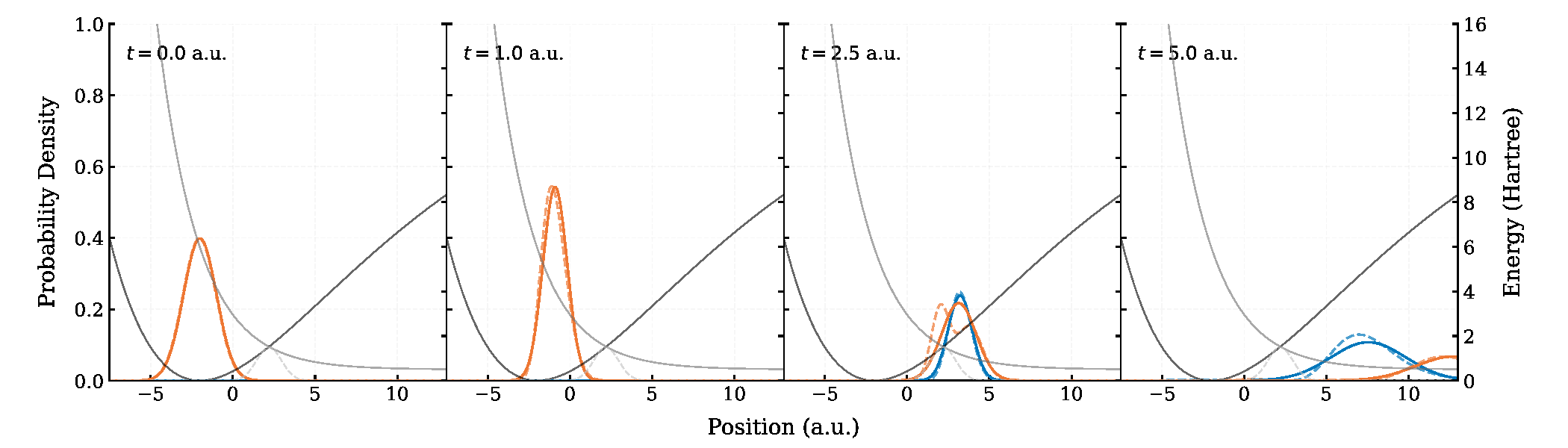}
\par\end{centering}
\caption{\label{fig:strip_shallow}Diabatic wavepackets time-evolution for
the deep Morse potential. Left axis for the probability density units
and right axis for the Hartree potential energy units. Gray for the
potential terms, blue for the lower $V_{11}$ state wavepacket and
orange for the upper dissociative $V_{22}$ state one. Colored solid
lines for NA-VTGWD, and colored dashed lines for the exact split-operator
quantum time evolution.}
\end{figure*}

We first investigate the dynamics of the deep Morse case. Fig. (\ref{fig:strip_shallow})
shows the snapshots of the wavepacket dynamics, as in Fig. (\ref{fig:dispHOs_wavepackets}).
The wavepacket is initially placed in the dissociative potential,
a regime where NA-VTGWD is highly accurate, as confirmed by the comparison
between the colored solid and dashed lines in Fig. (\ref{fig:strip_shallow}).
Importantly, this accuracy is preserved even after the wavepacket
crosses the non-adiabatic region, where most of the dissociative state
population is transferred to the bound Morse potential state.

\begin{figure}
\begin{centering}
\includegraphics[scale=0.3]{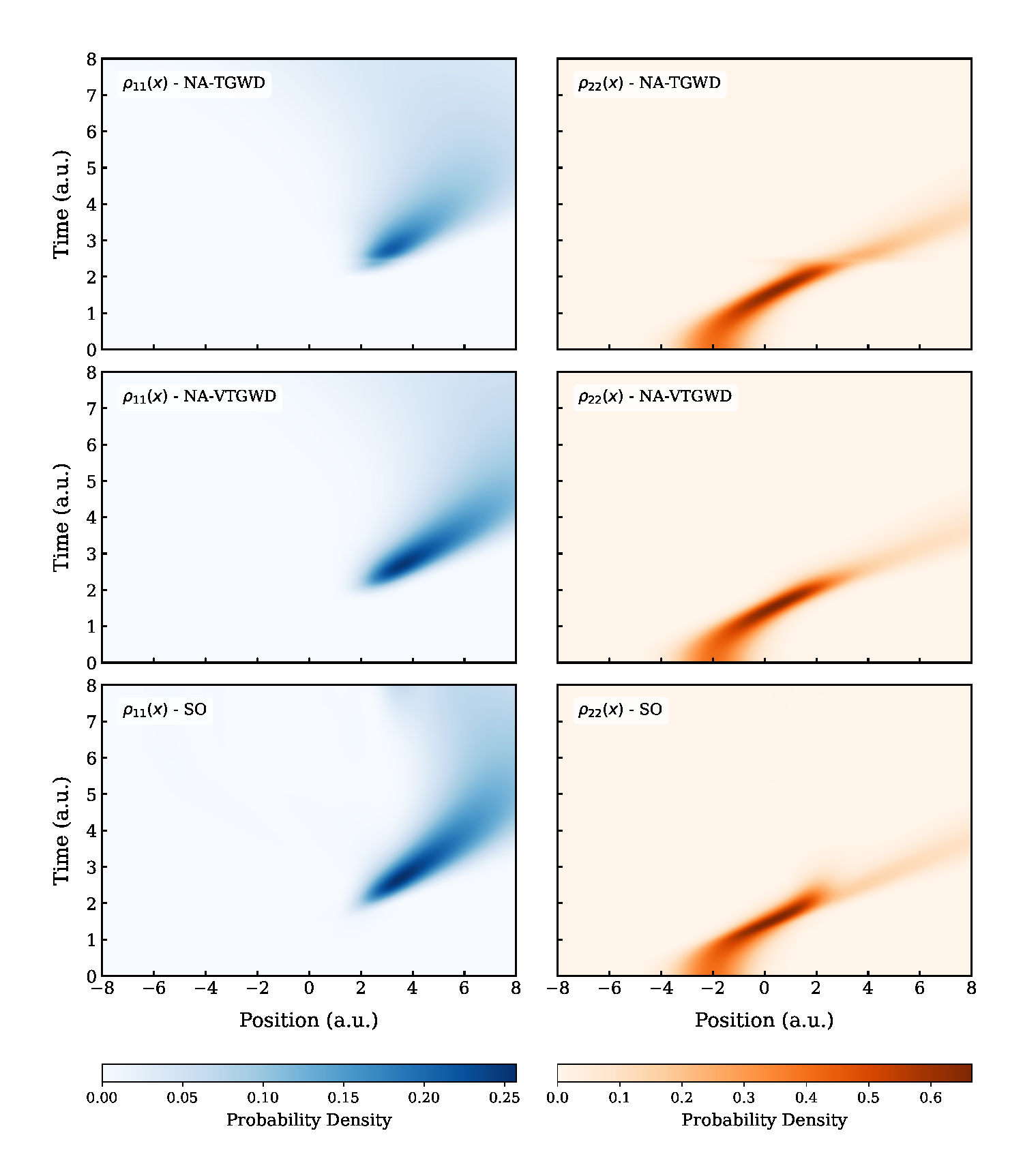}
\par\end{centering}
\caption{\label{fig:Heatmaps_shallow}Wavepacket heatmaps for the deep Morse
potential. Left for the lower $V_{11}$ electronic state. Right for
the upper dissociative $V_{22}$ one. Bottom panel is the exact quantum
evolution, middle one is NA-VTGWD, while upper panel is for NA-TGWD
approximation.}
\end{figure}

\begin{figure}
\begin{centering}
\includegraphics[scale=0.3]{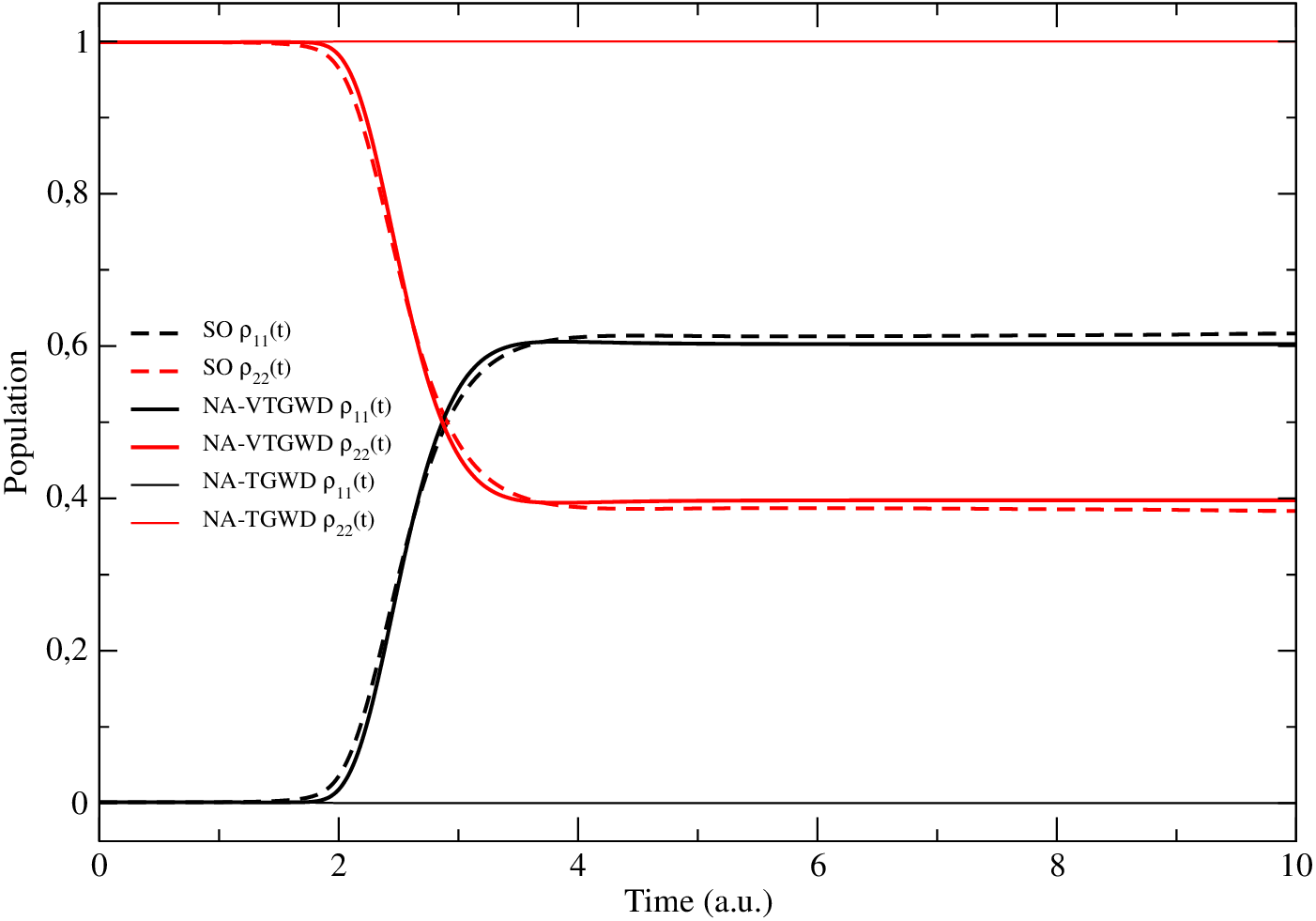}
\par\end{centering}
\caption{\label{fig:Population_deep}Diabatic populations time evolution for
the shallow Morse potential case.}
\end{figure}

A more comprehensive view is provided by the heatmaps in Fig. (\ref{fig:Heatmaps_shallow}),
where the dissociative potential population density is reported in
orange and the Morse one in blue. Once again, NA-VTGWD performs better
than NA-TGWD, which partially reproduces the population transfer.
NA-VTGWD is accurate throughout the entire dynamics. This is even
more evident when looking at the populations of Fig. (\ref{fig:Population_deep}),
where NA-VTGWD is almost exact, while NA-TGWD fails to fully capture
the transfer and is not able to recover the correct asymptotic population.
For a better understanding of the differences between the two methods,
we report in Fig.s S3 and S4 of the Supplementary Material the phase
space trajectories respectively for the NA-TGWD and the NA-VTGWD cases.
These phase space plots suggest that the variational motion trajectory
is smoother than the NA-TGWD case.

\begin{figure}
\begin{centering}
\includegraphics[scale=0.3]{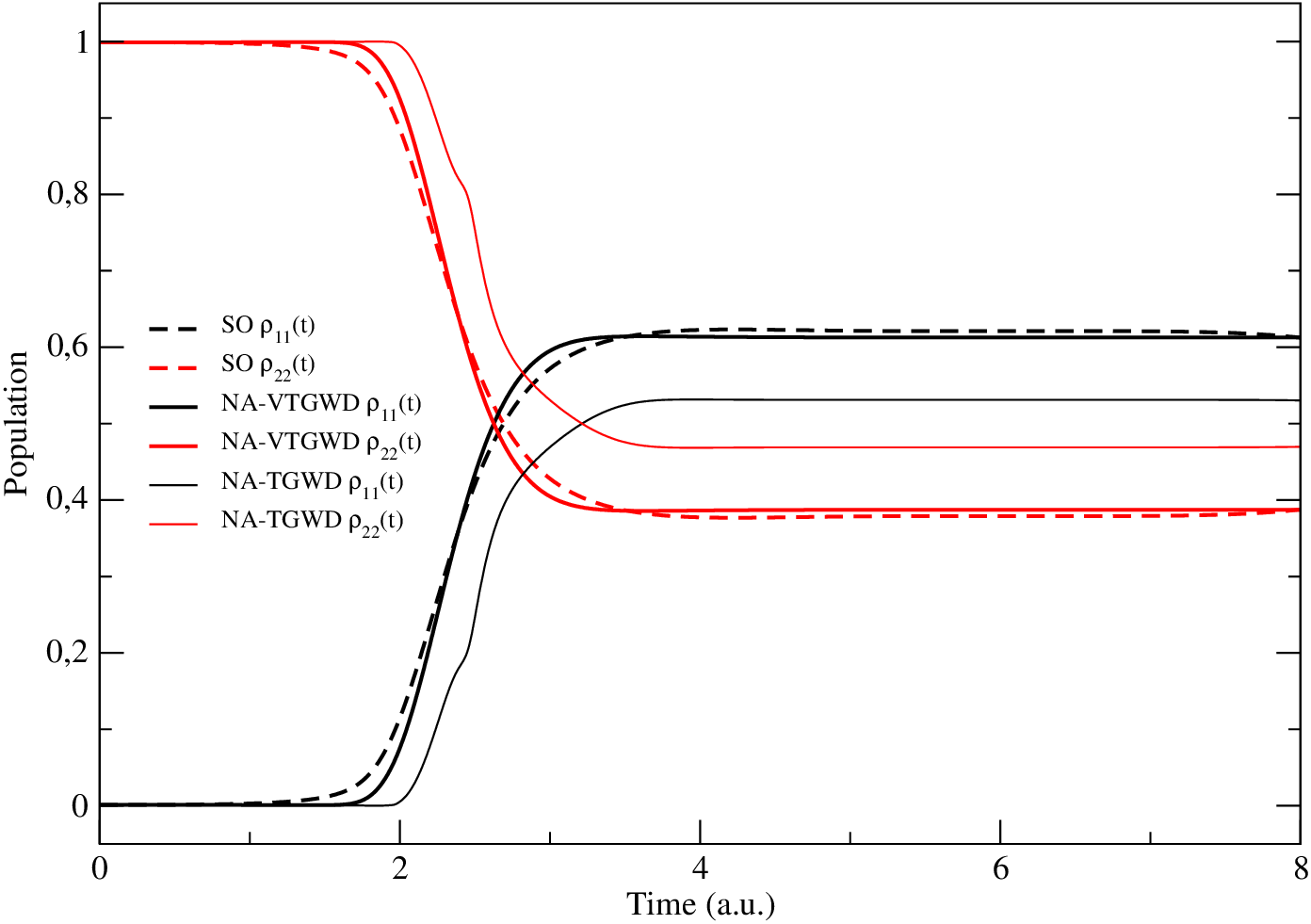}
\par\end{centering}
\caption{\label{fig:Population_shallow}Diabatic populations time evolution
for the deep Morse potential.}
\end{figure}

\begin{figure*}
\begin{centering}
\includegraphics[scale=0.5]{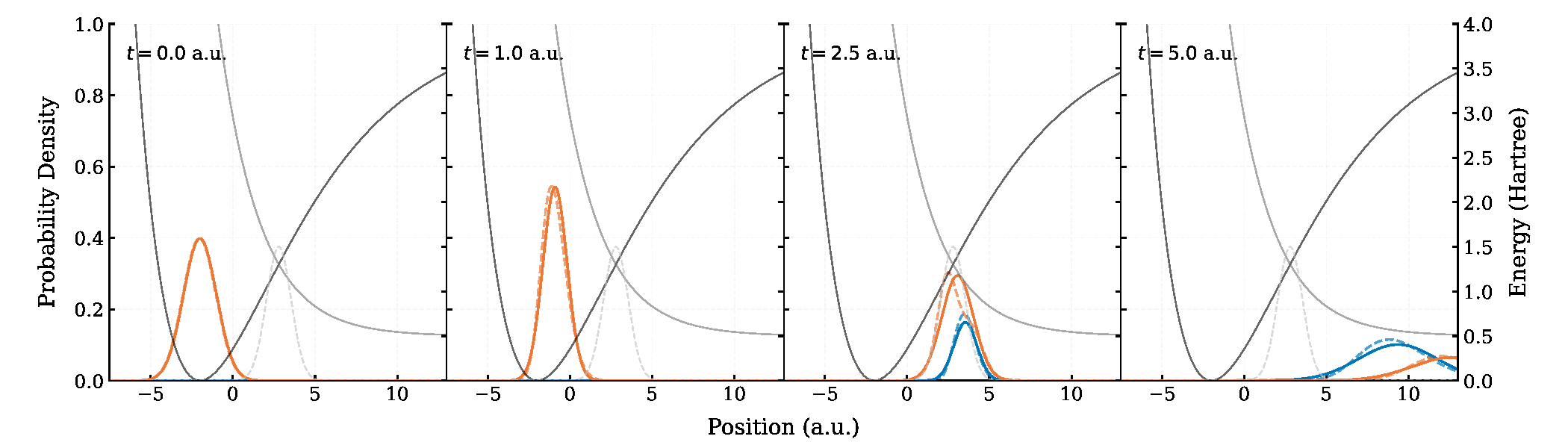}
\par\end{centering}
\caption{\label{fig:strip_deep}Diabatic wavepackets time-evolution for the
shallow Morse potential case (see units on the right axis). Gray for
the potential terms, blue for the lower $V_{11}$ state wavepacket
and orange for the upper dissociative $V_{22}$ state one. Colored
solid lines for NA-VTGWD, and colored dashed lines for the exact split-operator
quantum time evolution.}
\end{figure*}

We now turn to the shallow Morse potential case, where the performance
difference between the two methods becomes even more striking. Fig.
(\ref{fig:strip_deep}) shows the wavepacket evolution for the variational
method. The solid NA-VTGWD lines are barely distinguishable from the
dashed ones corresponding to the exact split-operator evolution.

\begin{figure}
\begin{centering}
\includegraphics[scale=0.3]{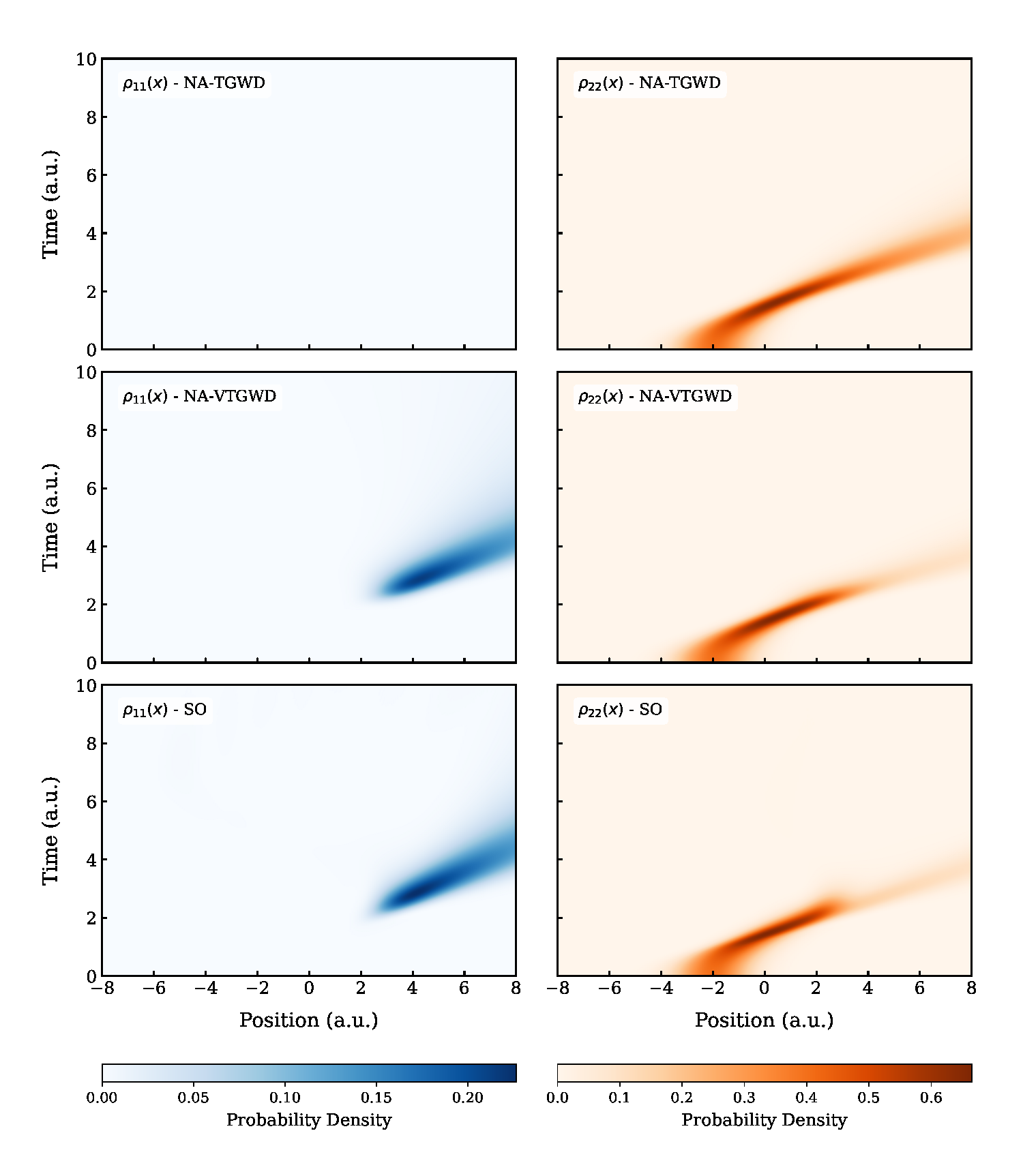}
\par\end{centering}
\caption{\label{fig:Heatmaps_deep}Wavepacket heatmaps for the shallow Morse
potential case. Left for the lower $V_{11}$ electronic state. Right
for the upper dissociative $V_{22}$ one. Bottom panel is the exact
quantum evolution, middle one is NA-VTGWD, while upper panel is for
NA-TGWD approximation.}
\end{figure}
 This accuracy is reflected in the heatmaps of Fig. (\ref{fig:Heatmaps_deep}).
In this case, the NA-TGWD approach fails completely to capture the
population transfer. This is due to the fact that from the very beginning
of the dynamics, the NA-TGWD wavepacket expansion is too localized
to contribute to the coupling between the wavepackets. Consequently,
the NA-TGWD evolution proceeds essentially as uncoupled adiabatic
dynamics. 
 These observations are futher confirmed by the population transfer
plot in Fig. (\ref{fig:Population_deep}), where the variational approach
is very accurate and the local NA-TGWD one completely misses the population
transfer. However, we suggest that NA-VTGWD accurately captures the
population inversion between the two electronic states in the cases
presented above because these types of photodissociation processes
involve only a single passage through the non-adiabatic coupling region.

\section{\label{sec:Discussion-and-Conclusions}Discussion and Conclusions}

A non-adiabatic semiclassical method is introduced and tested on model
diabatic potentials. The method has the advantage of being described
by a pair of coupled classical trajectories on separated surfaces,
i.e. classical trajectories follow classical mechanics and are not
forced to jump. This straightforward approach avoids any convergence
issues over an ensemble of classical trajectories and can be easily
implemented for on-the-fly ab initio non-adiabatic molecular dynamics.
Since the method is based on a single Gaussian description of the
non-adiabatic events, interference phenomena represented by multiple
wavepackets per potential energy surface cannot be described.

We find that the NA-TGWD approximation fails to fully capture population
transfer, whereas NA-VTGWD proves to be consistently accurate. This
comparison shows that nuclear quantum effects in non-adiabatic processes
can not be fully reproduced simply by a local Gaussian propagation.
Instead, a time-dependent variational approach, which effectively
accounts for non-local nuclear quantum effects, is necessary. This
aligns with previous findings, where Vanicek's group\citep{Vanicek_Rojia_VTGA_2023}
has already shown that in the adiabatic case, VTGWD can at least partially
recover tunneling, which is a challenging nuclear quantum effect to
reproduce. All NA-VTGWD space integrals have been performed analytically
and the method is expected to be computationally intensive if one
is forced to perform these integrals numerically. However, Vanicek's
group has also recently shown that if VTGWD integrands are approximated
up to the third order, the accuracy of the variational approach is
mostly retained.\citep{Moghaddasi_Vanicek_localcubicdynamics_2024}
This approximation is performed by applying a local cubic approximation
to the potential.\citep{ohsawa_TGAsymplectic_2013,pattanayak_extendedTGA_1994}
Also, on the numerical side, there is still room for improvement.
One can employ the Hagedorn formulation and exploit the symplectic
properties of the geometric integrators.\citep{hagedorn1998,vanicek_family_2023,Lauvergnat_Hagedorn_26}

In our formulation, we employ the diabatic framework instead of the
adiabatic one. One could think that this is a limitation because on-the-fly
ab initio molecular dynamics is performed in the adiabatic framework.
However, the non-adiabatic community is constantly developing diabatization
methods that allow one to perform, at least locally, quantum dynamics
using a diabatic potential.\citep{xie2026diabatization,BingGu_diabatization_26}
The diabatic potential is preferable to the adiabatic one because
non-adiabatic coupling terms can be singular in the adiabatic representation.\citep{mead1982conditions}
However, a rigorous diabatic representation does not exist because
of the non-removable residual derivative coupling and, for this reason,
the local diabatization is usually called quasi-diabatization.\citep{BinGu_diabaticCI_23,xie2026diabatization,BingGu_diabatization_26,Doriol_PhilTrans_22}

In conclusion, we have introduced a method capable of reproducing
non-adiabatic effects through classical mechanics by employing coupled
trajectories confined to separate potential energy surfaces

\section*{Supplementary Material}

Supplementary material contains the derivations of the main equations
and additional graphics about the coupled displaced harmonic oscillators
and the classical trajectory for the coupled deep Morse potentials.

\section*{Data Availability Statement}

The data that support the findings of this study are available from
the corresponding author upon reasonable request.
\begin{acknowledgments}
The authors thank Prof.s Riccardo Conte, Jiří Vaníček, Loïc Joubert-Doriol,
and Eli Pollak for useful discussions. M.C. thanks Università degli
Studi di Milano for funding under project PSR2025.
\end{acknowledgments}

\bibliographystyle{unsrt}
\bibliography{BiblioMAG26}

\end{document}